\documentclass[aps,prb,superscriptaddress,twocolumn,showpacs,amsmath,amssymb]{revtex4}
\usepackage{graphicx}% Include figure files
\usepackage{dcolumn}% Align table columns on decimal point
\usepackage{bm}% bold math
\begin{document}

\title{Study of HgBa$_2$CuO$_{4+\delta}$ by Angle-Resolved Photoemission Spectroscopy}
\author{W.S. Lee}
\affiliation {Department of Physics, Applied Physics, and Stanford
Synchrotron Radiation Laboratory, Stanford University, Stanford,
CA 94305}
\author{T. Yoshida}
\affiliation {Department of Physics, Applied Physics, and Stanford
Synchrotron Radiation Laboratory, Stanford University, Stanford,
CA 94305}
\author{W. Meevasana}
\affiliation {Department of Physics, Applied Physics, and Stanford
Synchrotron Radiation Laboratory, Stanford University, Stanford,
CA 94305}

\author{K.M. Shen}
\affiliation {Department of Physics, Applied Physics, and Stanford
Synchrotron Radiation Laboratory, Stanford University, Stanford,
CA 94305}

\author{D.H. Lu}
\affiliation {Department of Physics, Applied Physics, and Stanford
Synchrotron Radiation Laboratory, Stanford University, Stanford,
CA 94305}
\author{W.L. Yang}
\affiliation {Advanced Light Source, Lawrence Berkeley National
Lab, Berkeley, CA 94720}

\author{X.J. Zhou}
\affiliation {Advanced Light Source, Lawrence Berkeley National
Lab, Berkeley, CA 94720}

\author{X. Zhao} \affiliation {Department of Physics, Applied Physics, and
Stanford Synchrotron Radiation Laboratory, Stanford University,
Stanford, CA 94305}

\author{G. Yu}
\affiliation {Department of Physics, Applied Physics, and Stanford
Synchrotron Radiation Laboratory, Stanford University, Stanford,
CA 94305}

\author{Y. Cho}
\affiliation {Department of Physics, Applied Physics, and Stanford
Synchrotron Radiation Laboratory, Stanford University, Stanford,
CA 94305}

\author{M. Greven}
\affiliation {Department of Physics, Applied Physics, and Stanford
Synchrotron Radiation Laboratory, Stanford University, Stanford,
CA 94305}

\author{Z. Hussain}
\affiliation {Advanced Light Source, Lawrence Berkeley National
Lab, Berkeley, CA 94720}
\author{Z.-X. Shen}
\affiliation {Department of Physics, Applied Physics, and Stanford
Synchrotron Radiation Laboratory, Stanford University, Stanford,
CA 94305}

\date{\today}% It is always \today, today,

\begin{abstract}
We present the first angle-resolved photoemission measurement on
the single-layer Hg-based cuprate, HgBa$_2$CuO$_{4+\delta}$
(Hg1201). A quasi-particle peak in the spectrum and a kink in the
band dispersion around the diagonal of the Brillouin zone are
observed, whereas no structure is detected near the Brillouin zone
boundary. To search for a material-dependent trend among
hole-doped cuprates, including Hg1201, we use a tight-binding
model to fit their Fermi surfaces. We find a positive correlation
between the $T_{c,\mathrm{max}}$ and $t'/t$, consistent with
theoretical predictions.
\end{abstract}

\pacs{Valid PACS appear here}% PACS, the Physics and Astronomy
\maketitle

Empirically, it is known that the maximum transition temperature,
$T_{c,\mathrm {max}}$, varies strongly among different high-$T_c$
superconducting compounds \cite{Bi2212:inhomogeneity:Hiroshi}. For
example, the single-layer compounds, Bi$_2$Sr$_2$CuO$_{6+\delta}$
(Bi2201) and La$_{2-x}$Sr$_x$CuO$_4$ (LSCO), have a
$T_{c,\mathrm{max}}$ of $\sim$40 K, while HgBa$_2$CuO$_{4+\delta}$
(Hg1201) and Tl$_2$Ba$_2$CuO$_{6+\delta}$ (Tl2201) have a
$T_{c,\mathrm{max}}$ of 98 K and 93 K, respectively. This material
dependence has been related to disorder and crystal structure
\cite{Bi2212:inhomogeneity:Hiroshi,Material:Jogenson}. The latter
determines the hybridizations of the Cu orbitals with those of
other elements, resulting in different values of hopping integrals
in the effective single-band tight-binding Hamiltonian for
distinct compounds
\cite{LDA:Tc_Correlation,theoretical_calculation:apical oxygen}.
However, experimental studies which directly address this material
dependence of the band structure have not been extended to
compounds with the highest $T_{c,\mathrm{max}}$
\cite{IXS:LSCO,Tanaka:Bi2212}, due to the lack of sufficiently
large high-quality single crystals for the materials with higher
$T_{c,\mathrm{max}}$. Fortunately, through significantly improved
crystal growth techniques, sizable single crystals of Hg1201 have
recently become available \cite{Martin:crystal_growth}. Because
Hg1201 has the highest $T_{c,\mathrm{max}}$ among all single-layer
cuprates \cite{Material:First_Paper} and possesses a relatively
simple tetragonal crystal structure \cite{Material:Jogenson}, it
is an ideal system to complement this material dependence issue.
Until now, spectroscopy studies on this system have been limited
to angle-integrated photoemission \cite{AIPES:Hg1201} or to a
small momentum transfer region, such as Raman spectroscopy
\cite{Ramman:Hg1201}; the momentum space properties of the
electronic states, such as the band dispersions and the Fermi
surface (FS), are not yet available.

In this article, we report the first angle-resolved photoemission
spectroscopy (ARPES) results on nearly optimally-doped Hg1201
crystals (T$_c$=96 K). With the unique ability of ARPES to resolve
states in the momentum-energy space \cite{Group:Review}, we
observed a single sheet of the Fermi surface and a kink in the
dispersion along the diagonal of the zone (nodal direction). Our
analysis provides an opportunity to examine two universal
properties of the cuprates: (1) the positive correlation between
the effective single-band parameter, $t'/t$, and
$T_{c,\mathrm{max}}$; (2) the existence of an energy scale related
to the electron-boson coupling.

\begin{figure}
\includegraphics [ width = 3.0 in, clip]{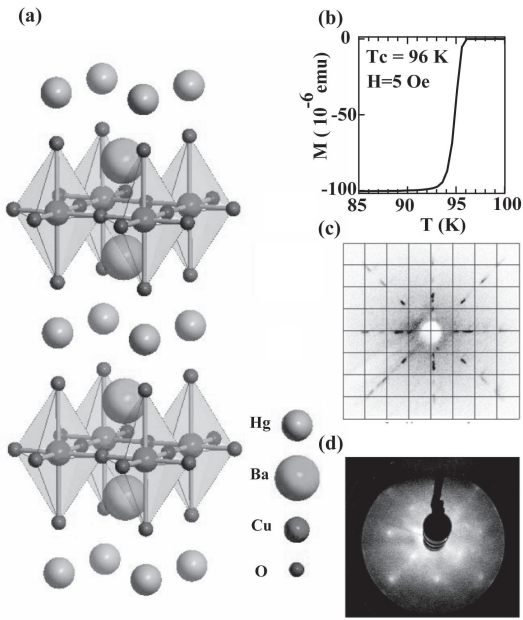}
\caption{\label{fig1:sample_properties} (a) Crystal structure of
Hg1201; (b) Zero field cooled magnetic moment curve of Hg1201; (c)
Laue pattern of Hg1201 representing the [001] surface, the a-b
plane; (d) LEED pattern of the cleaved surface.}
\end{figure}

Single crystals of Hg1201 were grown at Stanford University using
a significantly improved melt-growth method
\cite{Martin:crystal_growth}. Small pieces, with a typical size of
$0.7\times0.7\times0.5$ $\mathrm{mm}^3$, were selected for ARPES
measurements. The crystals were polished and cleaned using a
bromine solution (5\% Br + 95\% ethanol) until the surface was
shiny. The crystals were then annealed at 300$^\circ$C under
oxygen flow; finally, nearly optimally-doped crystals with a $T_c$
of 96 K and a transition width of less than 3 K were obtained, as
shown in Fig. \ref{fig1:sample_properties}(b). The bromine
treatment of the surface is applied again immediately before
mounting the samples to reduce the surface contact resistance. A
typical four-fold symmetric Laue pattern of the [001] surface is
shown in Fig. \ref{fig1:sample_properties}(c).

The ARPES experiments were performed at beamline 5-4 of the
Stanford Synchrotron Radiation Laboratory (SSRL) and beamline
10.0.1 of the Advanced Light Source (ALS) with SCIENTA SES200 and
R4000 electron analyzers, respectively. The energy resolution was
set at 20-25 meV. We used 19 eV photons at the SSRL and 55 eV
photons at the ALS to excite the electrons. The samples were
oriented \emph{ex situ} using Laue diffraction, then cleaved
\emph{in situ} and measured at a temperature of approximately 15K
under ultra high vacuum at a pressure lower than $4\times10^{-11}$
Torr. Perhaps due to the absence of a natural cleaving plane in
the crystal structure (Fig. \ref{fig1:sample_properties}(a)), the
cleaved surface of Hg1201 is generally rough, and the signal is
weaker compared to the other most studied cuprates, such as Bi2212
and LSCO. However, a LEED pattern can still be observed as shown
in Fig. \ref{fig1:sample_properties}(d), suggesting that a portion
of the surface layer still preserves good periodicity after
cleaving.

\begin{figure}
\includegraphics [ width = 3.25 in, clip]{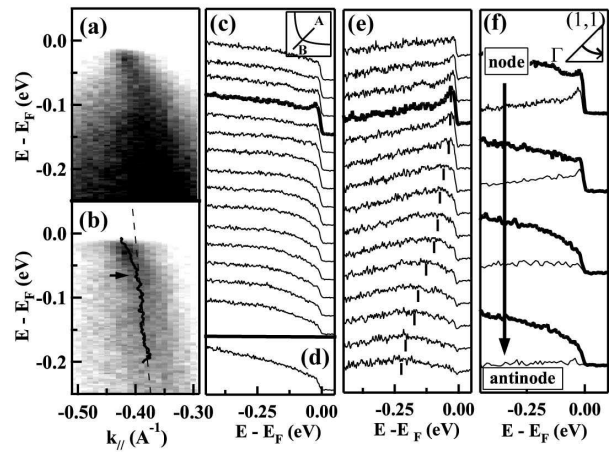}
\caption{\label{fig2:EDCs} (a) Grey scale intensity plot of the
raw ARPES spectrum along (0,0)-(1,1) direction; (b) Spectrum with
background [illustrated in (d)] subtracted from (a). The
MDC-derived band dispersion is illustrated by the solid curve; the
kink is indicated by the black arrow. (c),(e) EDC stack plots of
(a) and (b), respectively. Bars in (e) indicate the band
dispersion in EDCs. (d) The EDC of the background selected at a
position of $k\gg k_F$. (f) EDCs along the Fermi surface as
illustrated in the inset. The thicker curves represent the raw
data; the thinner curves illustrate the background removed data.}
\end{figure}

The grey scale intensity plot of the ARPES spectrum along the
nodal direction is illustrated in Fig. \ref{fig2:EDCs}(a), where a
band dispersion can be observed. Corresponding energy distribution
curves (EDCs) are plotted in Fig. \ref{fig2:EDCs}(c). A
quasi-particle-like peak can be observed near the Fermi crossing
point, $k_F$ (the thick curve in Fig. \ref{fig2:EDCs}(c)) and
rapidly fades into a step-like background. To better visualize the
band dispersion, the EDCs were first normalized between 500 meV
and 520 meV below the Fermi energy; we then subtracted the
normalized background (Fig. \ref{fig2:EDCs}(d)). The
background-removed spectra are shown in Fig. \ref{fig2:EDCs}(b)
and the corresponding EDCs are displayed in Fig.
\ref{fig2:EDCs}(e). Moving away from nodal region toward the the
antinodal region along Fermi surface, this quasi-particle peak
diminishes, as illustrated in Fig. \ref{fig2:EDCs}(f). Near the
antinodal region, the EDCs primarily consist of a step-like
background with no sharp quasi-particle-like peak in either the
raw or the background subtracted EDCs.

The appearance of this step-like background throughout the zone is
likely an angle-integrated spectrum due to the roughness of the
cleaved surface. We note that this background is gapped, with a
spectral weight suppression between $-30$ meV to E$_F$. This is
consistent with the angle-integrated photoemission spectrum
reported on the same material \cite{AIPES:Hg1201}, suggesting a
maximum superconducting gap size of approximately 30 meV or
larger.

The absence of a quasi-particle-like peak around the antinodal
region is, however, puzzling. Since the $T_c$ (96 K) of our
samples is comparable to the optimally-doped double-layer system
Bi2212, a sharp peak with high intensity in the antinodal region
as in Bi2212 is expected \cite{Bi2212:antinodal_peak}. This
antinodal peak, if it exists, should not be washed out by the
angle-integrated background, especially when a nodal
quasi-particle peak is present. Therefore, we suspect that the
absence of the antinodal peaks is either a result of broken
crystal symmetry at the surface, or an intrinsic feature of the
nearly optimally-doped Hg1201. For the latter, we note that it
might be similar to the case of LSCO in which there exists a
dichotomy electronic states in the nodal region and antinodal
region \cite{LSCO:XJZhou}; the antinodal peak may emerge only at
higher doping levels. A doping dependence study of Hg1201 will be
necessary to clarify this issue.

\begin{figure}
\includegraphics [ width = 3 in, clip]{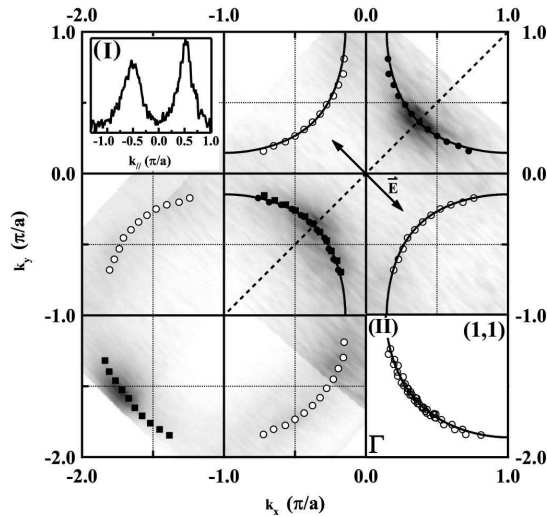}
\caption{\label{fig3:FermiSurface} Fermi surface of the Hg1201
generated by integrating the spectrum $\pm10$meV with respect to
E$_F$. The solid symbols represent the Fermi crossing points,
$k_F$, determined by the peak positions of the MDC near E$_F$. The
open symbols are symmetrized from the solid symbols. The solid
curve is the tight-binding FS as described in the text. The inset
(I) shows the MDC around E$_F$ along the dashed line. The inset
(II) illustrates the Fermi crossing points collected from three
sets of data which were taken at the SSRL and ALS. The fitted
tight binding FS is also shown.}
\end{figure}

In Fig. \ref{fig3:FermiSurface}, we show the FS of Hg1201, which
was taken using 55 eV photons at the ALS. As expected, a single
sheet of FS is observed. With our experimental setup, the spectral
weight in the second and fourth quadrants is strongly suppressed
due to the matrix element effect. The Fermi crossing points are
determined by fitting the peak positions of momentum distribution
curves (MDCs) near the Fermi energy, as shown in the inset of Fig.
\ref{fig3:FermiSurface}. The FS, illustrated by the solid symbols
in Fig. \ref{fig3:FermiSurface}, is extracted up to the antinodal
region where the MDC peaks disappear. We note that by collecting
data in symmetric quadrants on the same sample, the symmetry
requirement of the FS enables us to determine the position of the
Fermi crossing points more accurately. In particular, averaging
three data sets taken on different samples at the ALS and SSRL, we
determine the Fermi crossing point along the nodal direction,
$(k_n, k_n)$, to be $k_n=0.374\pm0.006$, in units of $\pi/a$.

Compared to other nearly optimally-doped cuprates, $k_n$ of Hg1201
is relatively small, close to that of materials with a higher
$T_{c,\mathrm{max}}$, as shown in Table \ref{table1:node}. The
variation in $k_n$ among nearly optimally-doped compounds, which
presumably possess a similar doping level, suggests a difference
of band structures for these systems. Setting aside strong
correlations, such as the electron-electron interaction and
renormalization effects from bosonic channel (e.g. phonons and
spin excitations), we use a tight-binding model to describe the
Fermi surface. Including up to the second nearest-neighbor hopping
term, the FS of the tight-binding model is given by
\begin{eqnarray}
0&=&\frac{\mu}{t}-2\left(\cos k_x+\cos k_y\right)
+4\frac{t'}{t}\cos k_x\cos k_y,
\label{eq:one}
\end{eqnarray}
where $t$ and $t'$ represent the nearest-neighbor and the second
nearest-neighbor hopping integral on Cu-O plane, respectively.
Based on Eq. \ref{eq:one}, we then preform a least-square fit for
the Fermi crossing points of various cuprates listed in Table
\ref{table1:node}. The Fermi crossing points of each compound are
collected from at least two sets of data, which are either
published results (e.g. Tl2201 data) or the data from our own
group. As an example, we show the Fermi crossing points of the
Hg1201 system in the inset (II) of Fig. \ref{fig3:FermiSurface},
which contains three data sets collected at the SSRL and ALS; the
FS obtained from the tight-binding fit is also plotted. Finally,
the ratio $t'/t$ obtained from our analysis is summarized in Table
\ref{table1:node}.

\begin{figure} [t]
\includegraphics [ width = 3.0 in, clip]{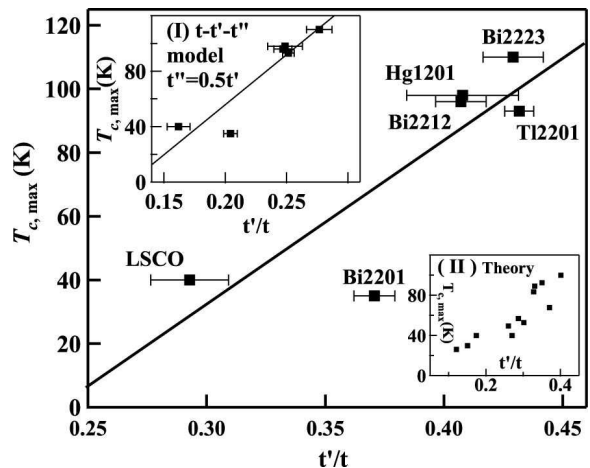}
\caption{\label{fig4:ratio} Positive correlation between
$T_{c,\mathrm{max}}$ and $t'/t$. Inset (I) is the results obtained
from $t-t'-t"$ model. Error bars are the 99\% confidence interval
of the fitted $t'/t$. Inset (II) is reproduced from Ref.
\cite{LDA:Tc_Correlation}. The lines in the figures are
guides-to-the eye, indicating the positive correlation between
$T_{c,\mathrm{max}}$ and $t'/t$.}
\end{figure}

To search for a material dependent trend, we plot $t'/t$ versus
the $T_{c,\mathrm{max}}$ of the corresponding compound in Fig.
\ref{fig4:ratio}; a positive correlation between the $t'/t$ and
the $T_{c,\mathrm{max}}$ is revealed. In the absence of FS data
for optimally doped Tl2201, the values of $t'/t$ obtained from the
overdoped Tl2201 ($T_c$ =20$\sim$30 K) \cite{Tl2201FS:Andrea,
Tl2210FS:Hussey} was used to represent those of the corresponding
optimally doped compounds. We note that the accuracy of $t'/t$ for
Tl2201 compound should be verified, once data for optimally-doped
crystals are available. We also tried to extend our tight-binding
model to include the third nearest-neighbor hopping term,
$-2\frac{t''}{t}(\cos 2k_x+\cos 2k_y)$. However, we find that the
fitting process does not converge very well; there exist many
possible sets of $t'/t$, $t''/t$, and $\mu/t$ yielding good fits
to the data. This is because the current FS data are not accurate
enough to uniquely determine the fitting parameters. To avoid this
difficulty, we set $t''$ to its leading-order value, $t'/2$
\cite{LDA:Tc_Correlation}, such that the fitting prcess is robust.
The fitted $t'/t$ is summarized in Table \ref{table1:node} and
plotted in the inset (I) of Fig. \ref{fig4:ratio}; similarly, the
positive correlation between $T_{c, \mathrm{max}}$ and $t'/t$ is
observed.

This correlation is consistent with theoretical calculations
\cite{LDA:Tc_Correlation,theoretical_calculation:apical oxygen},
although the correlation revealed from our analysis is less
monotonic than that from the theory (inset (II) of Fig.
\ref{fig4:ratio}). This is probably due to the limited accuracy of
our analysis and the strong correlation effects, which are not
captured by our analysis or band structure calculations.
Nevertheless, the qualitative agreement between our analysis and
the theories lends support to the theoretical proposal
\cite{LDA:Tc_Correlation,theoretical_calculation:apical oxygen}
that the crystal-structure along the $c-$axis, especially the
apical oxygen atom, can manifest itself in the effective
single-band structure of the $a-b$ plane. Furthermore, it has also
been shown by the mean-field calculations for the $t-J$ model that
a higher $t'/t$ ratio can enhance the stability of the
superconducting pairs \cite{tJmodel:TKLee,tJmodel:HQLin}. This is
also   consistent with the positive correlation between
$T_{c,\mathrm{max}}$ and $t'/t$. We remark that the correlation
observed here is probably only one factor of the material
dependence of $T_{c,\mathrm{max}}$, as there are other factors
that could also affect $T_{c,\mathrm{max}}$. For example, although
LSCO possesses a lower values of $T_{c,\mathrm{max}}$ and $t'/t$,
the competing orders in this system, such as stripes
\cite{LSCO:XJZhou}, could also suppress the $T_{c,\mathrm{max}}$.

\begin{table}
\caption{\label{table1:node} node position $k_n$, $T_c$,
$T_{c,\mathrm{max}}$, and $t'/t$ obtained from the tight-binding
analysis for various cuprates. Relevant references were also
included.}
\begin{ruledtabular}
\begin{tabular}{cccccccc}
 &$k_n\footnotemark[1]$&$T_c(K)$&$T_{c,\mathrm{max}}$&$t'/t$\footnotemark[2]&$t'/t$\footnotemark[3]&reference\\
\hline
Hg1201&0.374&96&98&0.408&0.249&\cite{Tc_doping:Hg1201}\\
Bi2223&0.367&110&110&0.429&0.277& \cite{Bi2212:inhomogeneity:Hiroshi}\\
Bi2212&0.381&96&96&0.407&0.247&\cite{Bi2212:inhomogeneity:Hiroshi}\\
Tl2201&0.357&20$\sim$30&93&0.433&0.251&\cite{Tl2201FS:Andrea,Tl2210FS:Hussey}\\
LSCO&0.400&40&40&0.293&0.162&\cite{LSCO:XJZhou,LSCO:Ino}\\
Bi2201&0.397&35&35&0.371&0.204&\cite{Bi2212:inhomogeneity:Hiroshi}\\
\end{tabular}
\end{ruledtabular}
\footnotetext[1] {For Tl2201, $k_n$ was obtained by digitizing the
FS shown in the Ref. \cite{Tl2201FS:Andrea}; others values were
obtained directly from our group's data.} \footnotetext[2]
{Obtained by fitting FS to the $t-t'$ model.} \footnotetext[3]
{Obtained by fitting FS to the $t-t'-t''$ model with the constrain
$t''=0.5t'$.}
\end{table}

Our data also confirm another universal property of the cuprates:
the existence of an energy scale which is related to the
electron-boson coupling. In Fig. \ref{fig2:EDCs}(b), we superpose
the MDC-derived band dispersion on the grey scale intensity plot
of the spectrum. A kink in the dispersion between 60 meV to 80 meV
is observed. This feature has also been observed in other cuprates
\cite{CollectiveMode:Lazara}, suggesting that it is a universal
feature in cuprates. This dispersion kink has recently attracted a
lot of attention in the high $T_c$ community and has been thought
to be a signature of electrons coupled to some bosons, whose
origins are still strongly debated
\cite{CollectiveMode:Eschrig,CollectiveMode:Lazara}. Possible
candidates in the case of Hg1021 system include the following
modes: (1) the A$_{1g}$ 590 cm$^{-1}$ phonon mode observed in the
Raman spectrum, which shows a superconductivity-induced anomaly
\cite{Ramman:Hg1201}; (2) the bond stretching phonon, which shows
an unusual softening at a similar energy
\cite{IXS:Hg1201:strechingPhonon}; (3) a mode of magnetic origin
\cite{Hg1201:Anti-Ferro}. More detailed studies are needed to
identify the origin of the bosonic mode(s) in Hg1201.

In conclusion, we report the first ARPES results of the nearly
optimally-doped Hg1201 crystals. A tight-binding analysis on the
FS of various optimally-doped cuprates is also illustrated, which
suggests a positive correlation between $t'/t$ and the
$T_{c,\mathrm{max}}$. However, a more complete analysis, which
takes the strong correlation effects into account, is necessary to
gain deeper insight into the material dependence issue. Regarding
the universality of the electron-boson coupling in cuprates,
further studies covering a broader range of materials and doping
levels are required for a comprehensive understanding.

W.S. Lee thanks S. Maekawa, T. Devereaux, F. Baumberger, and O.K.
Andersen for useful discussions. SSRL is operated by the DOE
Office of Basic Energy Science, Division of Chemical Science and
Material Science under contract DE-AC03-765F00515. The ARPES
measurements at Stanford were also supported by NSF DMR-0304981,
ONR Grant No. N00014-0501-0127. The crystal growth was supported
by DOE Contracts No. DE-FG03-99ER45773 and DE-AC03-76F00515, and
by NFS DMR 9985067.

\end{document}